\documentclass[conference]{IEEEtran}
\IEEEoverridecommandlockouts
\usepackage{cite}
\usepackage{amsmath,amssymb,amsfonts}
\usepackage{algorithmic}
\usepackage{graphicx}
\usepackage{float}
\usepackage{tikz}
\usepackage{textcomp}
\usepackage{xcolor}
\def\BibTeX{{\rm B\kern-.05em{\sc i\kern-.025em b}\kern-.08em
    T\kern-.1667em\lower.7ex\hbox{E}\kern-.125emX}}
    
\newcommand\copyrighttext{%
  \footnotesize \textcopyright 2021 IEEE. Personal use of this material is permitted.
  Permission from IEEE must be obtained for all other uses, in any current or future
  media, including reprinting/republishing this material for advertising or promotional
  purposes, creating new collective works, for resale or redistribution to servers or
  lists, or reuse of any copyrighted component of this work in other works.}
\newcommand\copyrightnotice{%
\begin{tikzpicture}[remember picture,overlay]
\node[anchor=south,yshift=10pt] at (current page.south) {\fbox{\parbox{\dimexpr\textwidth-\fboxsep-\fboxrule\relax}{\copyrighttext}}};
\end{tikzpicture}%
}

\begin{document}
\copyrightnotice
\title{Regression with Deep Learning for Sensor Performance Optimization}

\author{
    \IEEEauthorblockN{Ruthvik Vaila\IEEEauthorrefmark{1} \IEEEauthorrefmark{2}, Denver Lloyd\IEEEauthorrefmark{2}, Kevin Tetz\IEEEauthorrefmark{2}}
    \IEEEauthorblockA{\IEEEauthorrefmark{1}Boise State University
    \\\{ruthvikvaila\}@u.boisestate.edu}
    \IEEEauthorblockA{\IEEEauthorrefmark{2}ON Semiconductor    
	\\\{Ruthvik.Vaila, Denver.Lloyd, Kevin.Tetz\}@onsemi.com}
}

\maketitle

\begin{abstract}
 Neural networks with at least two hidden layers are called deep networks \cite{Nielsen}. Recent developments in AI and computer programming in general has led to development of tools such as Tensorflow, Keras, NumPy etc. making it easier to model and draw conclusions from data. In this work we re-approach non-linear regression with deep learning enabled by Keras and Tensorflow. In particular, we use deep learning to parametrize a non-linear multivariate relationship between inputs and outputs of an industrial sensor with an intent to optimize the sensor performance based on selected key metrics. 
\end{abstract}

\vspace*{-0.5cm}
\section{Introduction}
\begin{figure}
\centering
\includegraphics[
height=2.2in,
width=3.2in
]%
{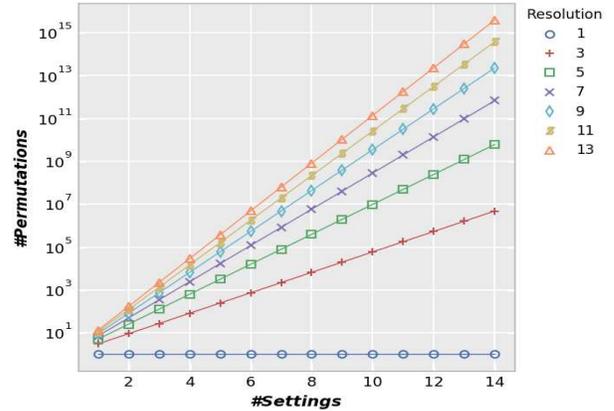}%
\caption{Resolution indicates number of values a particular setting can assume.}%
\label{n_settingsvspermuta}%
\end{figure}

\vspace*{-0.2cm}
Deep learning has been used in plethora of applications like autonomous driving, cancer prediction, low power object recognition etc \cite{vaila2019} \cite{vaila2019a} \cite{vishal2018}. In particular, neural networks as a regression tool have been used in applications like, time series learning \cite{Schmidhuber}, stock prediction \cite{Refenes}, pose estimation in computer vision \cite{Lathuil}, cost predictions \cite{SMITH} etc. Traditionally, linear regression with linear or non-linear coefficients has been used for modeling where real valued outputs are required. Universal approximation theorem states that a feed forward neural network with at least one hidden layer  can approximate a continuous function of $\mathbb{R}^{n}$ \cite{Hornik}. Neural networks use stochastic gradient descent (SGD) \cite{Lecun} to achieve an acceptable local minima that optimizes the output loss function. Many industrial sensors require fine tuning of the input settings to attain a desired output. Figure \ref{n_settingsvspermuta} shows that the number of experiments to be conducted increases by orders of magnitude with increase in resolution and number of inputs to a sensor. In this work, we employed deep learning to model the relationship between inputs and outputs of a sensor that were collected at set intervals. Once a satisfactory model was achieved, we used the model to interpolate the outputs for any input  combinations of the sensor that are within an allowed range for that input. Using appropriate optimization criteria we showed that we arrived at an input setting that maximized or minimized required outputs of the given sensor. 
\vspace{-0.3cm}
\section{Background}

We used a quadratic cost function on the output layer. Cost, C is calculated by \cite{Nielsen}
\vspace{-0.2cm}
\[
C \triangleq \frac{\left\| y-a^{L} \right\|^{2}}{2} \tag{BP1}
\]

where $y$, $a^{L}, a^l (=\sigma(z^l) ) $ are the label vector, the output layer activation vector, and an activation vector of the $l^{th}$  layer respectively. Weights ($ w^l$) and biases ($ b^l$) are modified to decrease C. Error vector on the output layer is given by
\[
\delta^L = -(y - a^{L})\odot \sigma'(z^L)  \tag{BP2} \label{bp1b}
\]

where $\sigma,  z^l (=w^lz^{l-1}+b^l)$ represent a chosen activation function of the neurons and net input to the $l^{th}$  layer respectively. Error vector for internal layers is given by
\[
\delta^l = ((w^{l+1})^T \delta^{l+1}) \odot \sigma'(z^l) \tag{BP3}
\]
where $\odot$ denotes element wise multiplication. Updates to biases and weights of a layer $l$ are given by
\[
\frac{\partial C}{\partial b^l} = \delta^l \tag{BP4}
\]
\vspace{-0.2cm}
\[
\frac{\partial C}{\partial w^l} = \delta^l a^{(l-1)T} \tag{BP5}
\]

Final weight update equation for layer $l$ is given by

\[
  w^l =   w^l -\eta \frac{\partial
    C}{\partial w^l}. \tag{BP6}
\]

similarly, biases of layer $l$ are updated according to

\[
 b^l = b^l - \eta \frac{\partial C}{\partial b^l} \tag{BP7}
\]

where $\eta$ is the learning rate which was set to $0.0005$.

\section{Dataset}

\begin{figure}
\centering
\includegraphics[
height=3.5in,
width=3.5in
]%
{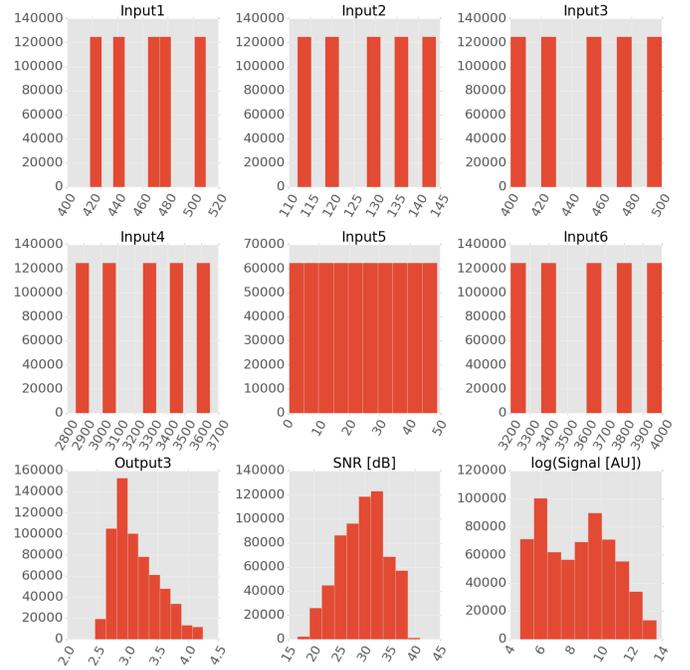}%
\caption{Histogram of all the numerical inputs and outputs.}%
\label{histograms}%
\end{figure}

Throughout the paper, we use the sensor data obtained from ON Semiconductor. Given sensor has seven inputs and three outputs, six of the inputs are numerical and seventh input is categorical and can take four possible values.
Histograms of all the numerical inputs and  outputs are shown in Figure \ref{histograms}. Categorical variable is not shown in Figure \ref{histograms}. Each of the inputs(1 to 4 and 6) assume five different values therefore we have a total of $5^{5}(3125)$ possible combinations. For each of the possible combinations, Input5 was swept from $0-49$. As there are four categorical variables, each of the input setting combinations yields a table (DataFrame) of $50*4(=200)$ rows. As there are $3125$ possible setting combinations the final dataset contains $3125*200 (=625000)$ rows and each row  is applied as an input to the sensor resulting in three outputs consisting of Signal, SNR and Output3. Therefore, the input to the neural network is $\in  \mathbb{R}^{625000 \times 10}$ and the output is $\in \mathbb{R}^{625000 \times 3}$.

\begin{table}[th]%
\caption{Concerned sensor of this work was presented with all the combinations of Input1, Input2, Input3, Input4, Input6 values given in the table. For each of the combination, Input5 was swept from $0-49$ obtaining a single Signal [AU] vs SNR [dB] curve. Note that the resolution of inputs for which outputs were recorded is 22, 8, 25, 200 and 200 for Inputs 1, 2, 3, 4 and 6 respectively.}
\centering
\begin{tabular}
[c]{|c|c|c|c|c|}\hline
\textbf{Input1} & \textbf{Input2} & \textbf{Input3} &
\textbf{Input4} & \textbf{Input6} \\\hline

418 & 112 & 400 & 2850 & 3200\\\hline
441 & 120& 425 & 3050 & 3400\\\hline
464 & 128& 450 & 3250 & 3600\\\hline
478 & 136& 475 & 3450 & 3600\\\hline
510 & 144& 500 & 3650 & 4000\\\hline
\end{tabular}
\label{settings_table}
\end{table}

\section{Neural network}
\begin{figure}
\centering
\includegraphics[
height=2.2in,
width=3.5in
]%
{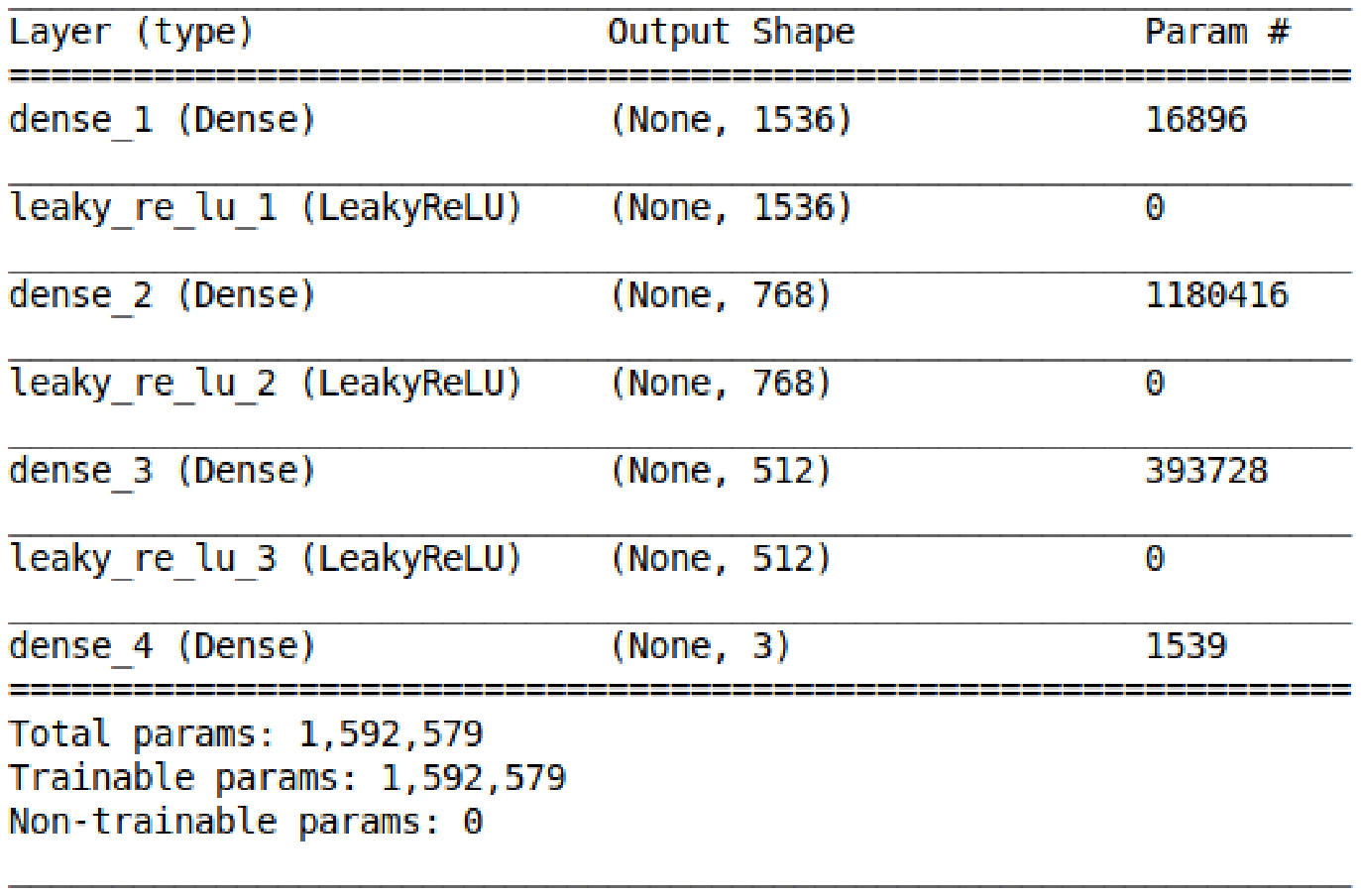}%
\caption{Keras sumary of the final neural network that was used to model the data.}%
\label{modelsummary}%
\end{figure}

A neural network with three hidden layers, mean squared error (MSE) cost function and a leaky ReLU activation function ($\sigma$) was chosen. Our network has $10$ input and $3$ output neurons which are determined by the dataset. Training was performed using Keras \cite{Chollet} with Tensorflow \cite{Abadi} back end and Adam was the chosen optimizer . Network's Keras summary is given in Figure \ref{modelsummary}. 
\subsection{Data pre-processing}
The Signal vs SNR relation of the data from the concerned sensor is approximately log linear for initial Signal values, it is shown in the Figure \ref{single_settings_snrvssignal}. The Signal [AU] column of the dataframe was log transformed and all the inputs to the neural network were normalized  by dividing an input with the maximum value it could assume. Therefore, all the inputs to the neural network are in between $0$ and $1$. Outputs were similarly normalized. All the data were converted to dataframes using Pandas \cite{McKinney}.

\subsection{Modeling}
Data were split into training ($81 \%$ ), validation ($9 \%$ ) and testing ($10 \%$ ). Our network was trained for $100$ epochs and learning rate was reduced by a factor of two for every consecutive five epochs if the validation error did not decrease. Batch size was set to $20$ and we report the mean square error (MSE) to be $4\times10^{-5}$ on the validation set.

\subsection{Prediction and Evaluation}
\begin{figure}[H]
\centering
\includegraphics[
height=1.75in,
width=2.15in
]%
{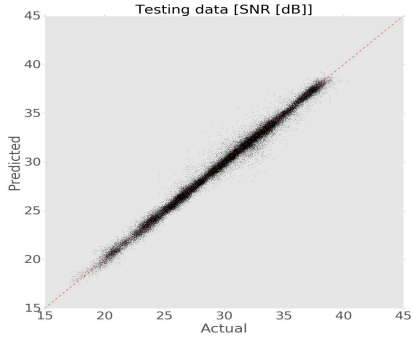}%
\caption{Actual vs Predicted plot for SNR [dB] in the testing dataset. Goodness of fit ($R^{2}$) was found to be $0.990$}%
\label{snr_test_predvsactual}%
\end{figure}

Once the model was trained, training data, testing data and validation data were passed through the network to obtain the predictions for the required outputs (SNR[dB], Signal [AU], Output3). Note that the model/network has not 'seen' the testing data directly and validation data was 'seen' indirectly in that it was used to optimize for the learning rate. Figure \ref{snr_test_predvsactual} shows the Actual vs Predicted plot for SNR[dB] in testing data and it is a linear plot indicating that the model was successful in predicting the SNR[dB] values for unseen data.

\begin{figure}[H]
\centering
\includegraphics[
height=1.75in,
width=2.15in
]%
{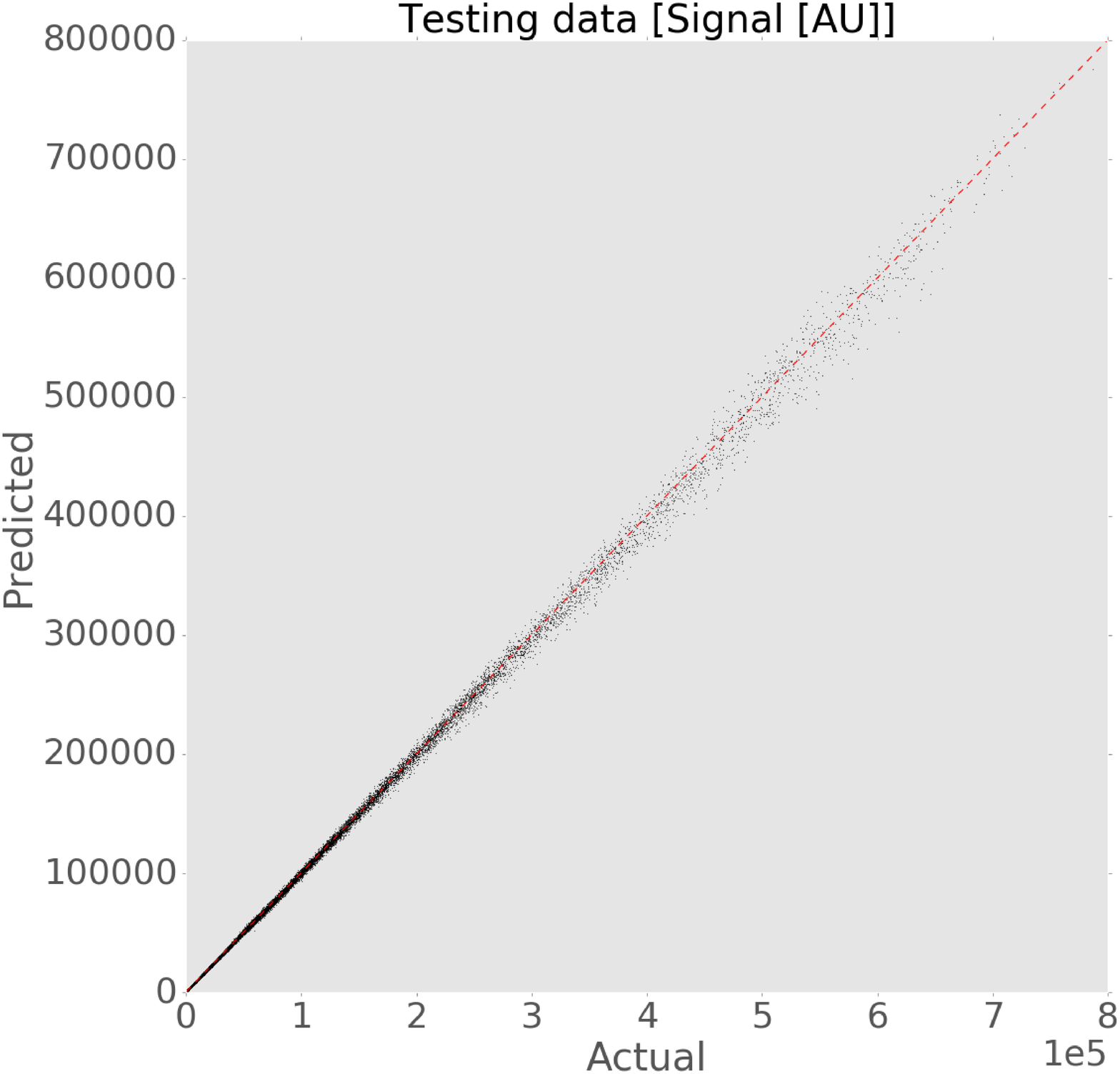}%
\caption{Actual vs Predicted plot for Signal [AU] in the testing dataset. Goodness of fit ($R^{2}$) was found to be $0.999$ }%
\label{signal_test_predvsactual}%
\end{figure}

\begin{figure}[H]
\centering
\includegraphics[
height=1.5in,
width=2.0in
]%
{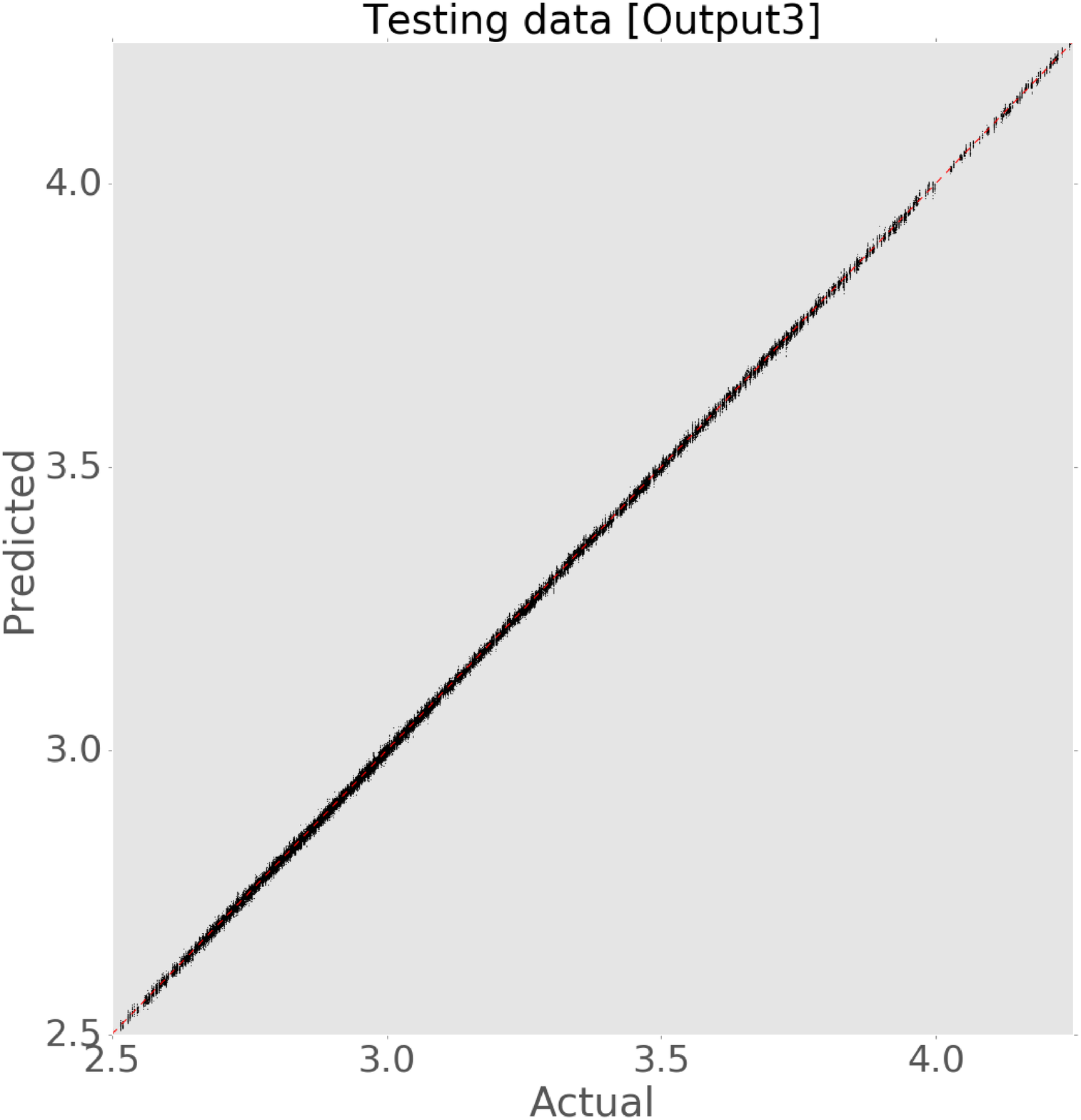}%
\caption{Actual vs Predicted plot for Output3 in the testing dataset.  Goodness of fit ($R^{2}$) was found to be $0.999$ }%
\label{output3_test_predvsactual}%
\end{figure}

 Figures \ref{signal_test_predvsactual}, \ref{output3_test_predvsactual} show Actual vs Predicted plots of Signal [AU], Output3 for testing datasets.

\section{Optimization}
\begin{figure}[H]
\centering
\includegraphics[
height=1.5in,
width=3.5in
]%
{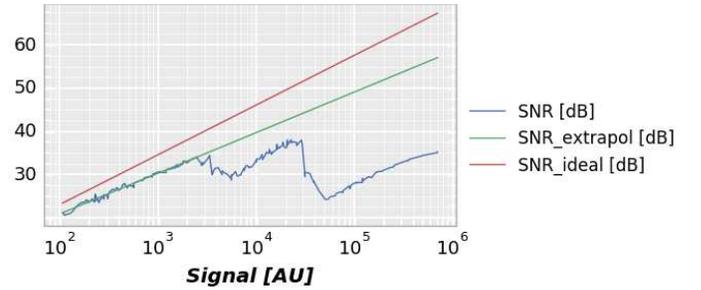}%
\caption{Plot of Signal [AU] vs  SNR [dB].}
\label{single_settings_snrvssignal}%
\end{figure}

The goal of the optimization process is to obtain a settings combination (of Input1, Input2, Input3, Input4 and Input6) that results in a Signal [AU] vs SNR [dB] curve that is closest to the ideal one and minimize the value of Output3. For the sensor under consideration, the ideal $SNR [dB] = 10\log_{10}(\sqrt{Signal [AU]}). $ Each of the settings combinations (of  Input1, Input2, Input3, Input4 and Input6) results in a dataframe of $200$ rows because Input6 is swept from $0-49$ and the categorical variable assumes four different categories and each of these dataframes yields a single Signal [AU] vs SNR [dB] curve. Note that Signal [AU], SNR [dB] and Output3 are the outputs of the trained neural network. The trained model was used to predict Signal [AU] vs SNR [dB] plots for a large number ( $\approx 12\times 10^{6} $) of interpolated settings combinations within the domains of all the input settings, to that end we increased the resolution of the numerical inputs listed in Table \ref{settings_table}. Similar to the original dataset, each of the interpolated input settings combinations also yields a single Signal [AU] vs SNR [dB] curve. Shown in Figure \ref{single_settings_snrvssignal} is a Signal [AU] vs SNR [dB] curve for a randomly chosen interpolated input settings, green and blue colors indicate ideal and predicted Signal [AU] vs SNR [dB] relationships. In this case,  Input1, Input2, Input3, Input4 and Input6 happened to be 418, 112, 400, 2850 and 3200 respectively and the value of Output3 is $2.9365$. The green colored line indicates the fitted line of Signal [AU] with SNR [dB] for Signal [AU] values that are less than $2\times 10^3$ .

\begin{figure}[H]
\hspace*{-2.5cm}      
\centering
\includegraphics[
height=1.25in,
width=1.8in
]%
{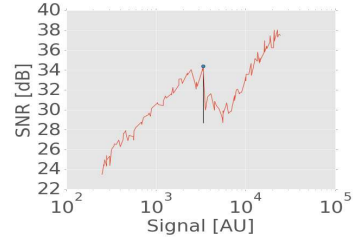}%
\caption{Zoomed plot of Signal [AU] vs  SNR [dB] in the  interval $\approx 3\times 10^3 - 10^{4}$ AU from the sensor for a single settings combination. Recorded prominence (SNR[dB] drop) value for this settings combination was $\approx 5.77$ dB. }
\label{zoomed_prominence}%
\end{figure}

 Blue curve in Figure \ref{single_settings_snrvssignal} shows a linear relationship until Signal [AU] reaches $\approx 3\times 10^{3} $ AU. Ideally, we expect this behavior to continue for the rest of the Signal values. A sudden dip of $\approx 5$ dB is noticeable when the Signal [AU] value is in the range, $\approx 3\times 10^3 - 10^{4}$ AU. Since it is highly unlikely to achieve an ideal performance, we set a few criteria to choose a particular settings combination that could give the smallest dip in the SNR value at the interval $\approx 3\times 10^3 - 10^{4}$ AU and a Signal [AU] vs SNR [dB] curve that is closest to the ideal Signal [AU] vs SNR [dB] curve. The best interpolated input combination was filtered by applying different criterion described below.  Lower values are preferred for all the criteria.

\begin{figure}[H]
\centering
\includegraphics[
height=1.25in,
width=3.5in
]%
{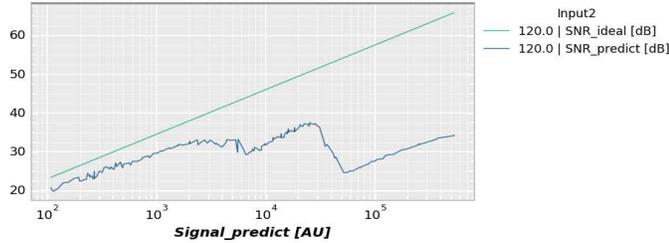}%
\caption{Plot of Signal [AU] vs  SNR [dB] when all the criteria were considered. Input1, Input2, Input3, Input4, Input6 were found to be $430, 120, 485, 2900, 3525$ respectively.}
\label{single_signalvssnr_op3_optim}%
\end{figure}

\begin{itemize}
  \setlength{\itemindent}{0em}
  \item [$\bullet$] \textbf{MAE between ideal and predicted (criterion 1):} Mean Absolute Error (MAE) was calculated for each of the input combinations and serial numbers of each of the dataframes (a single settings combination)  was ordered in an ascending order of the calculated MAEs.
  
  \item  [$\bullet$] \textbf{Prominence of SNR dip (criterion 2):}  Serial numbers of  each of the dataframes (a single input settings combination)  was ordered in an ascending order of the calculated dip in SNR [dB] value at $\approx 3\times 10^3 - 10^{4}$ AU.
  
  \item [$\bullet$] \textbf{MAE between fitted line and predicted (criterion 3):} Serial numbers of each of the dataframes (of a single input settings combination)  was ordered in an ascending order of the calculated for MAE between fitted green line and predicted blue curve of Figure \ref{single_settings_snrvssignal}. Green line was fitted for Signal [AU] vs SNR [dB] upto  $\approx 3\times 10^3 - 10^{4}$ AU and extrapolated for the rest of the Signal [AU] values. 

 \item [$\bullet$] \textbf{Least value for Output3 (criterion 4):} Serial numbers of  each of the dataframes (a single input settings combination)  was ordered in an ascending order of the calculated Output3 value. 
\end{itemize}
\vspace*{-0.15cm}
The first index among the intersection of all the indices obtained from the above steps gives the optimal input setting combination with a Signal [DN] vs SNR [dB] curve that meets all the above criteria. Figure \ref{single_signalvssnr_op3_optim} shows the optimized curve.
\vspace*{-0.35cm}

\vspace*{-0.45cm}

\begin{table}[th]%
\caption{criteria values when optimized for both SNR [dB] and Output3.}
\centering
\begin{tabular}
[c]{|c|c|c|c|}\hline
\textbf{criterion 1} & \textbf{criterion 2} & \textbf{criterion 3} & \textbf{criterion 4} \\\hline
1167.50 & 3.9  & 384.73  & 2.64\\\hline
\end{tabular}
\label{optim3_table}
\end{table}

\begin{figure}[H]
\centering
\includegraphics[
height=1.25in,
width=3.5in
]%
{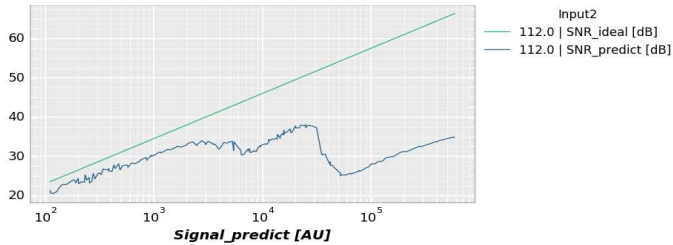}%
\caption{Plot of Signal [AU] vs  SNR [dB] when criterion 4 was ignored. Input1, Input2, Input3, Input4, Input6 were found to be $426, 112, 495, 3000, 3600$ respectively. }
\label{single_signalvssnr_op3_not_optim}%
\end{figure}

Table \ref{optim3_table} shows the numerical values of different criteria used in the optimization process. If criterion 4 was excluded from the optimization criteria (i.e., Signal [AU] vs SNR [dB] curve not optimized for Output3) then the Signal [AU] vs SNR [dB] is shown in Figure \ref{single_signalvssnr_op3_not_optim} and corresponding values of criteria are shown in Table \ref{not_optim3_table}. 

\begin{table}[th]%
\caption{criteria values when optimized only for SNR [dB].}
\centering
\begin{tabular}
[c]{|c|c|c|c|}\hline
\textbf{criterion 1} & \textbf{criterion 2} & \textbf{criterion 3} & \textbf{criterion 4}  \\\hline
1043.09 & 3.68 & 372.75  & 2.88 \\\hline
\end{tabular}
\label{not_optim3_table}
\end{table}

Figure \ref{single_signalvssnr_op3_not_optim} was obtained by optimizing for only Signal [AU] vs SNR [dB] curve. Hence, criterion 4 of the Table \ref{not_optim3_table} shows higher value than that of criterion 4 in the Table \ref{optim3_table}.

\section{Conclusion}
We have shown that deep neural networks can be successfully used for black box modeling of industrial sensors and the obtained model can be used to significantly speedup and improve the sensor performance optimization.
\section{Acknowledgments}
 Kevin Tetz proposed the idea and collected the sensor data, Ruthvik Vaila pre-processed the sensor data and setup the Machine Learning and Optimization pipelines, Denver Lloyd assisted in coding and discussions regarding the project. We thank the management at ON Semiconductor for providing us an opportunity to publish the work.
\bibliographystyle{ieeetr}
\bibliography{wmed2020}

\end{document}